\documentclass[openacc]{rstransa}

\usepackage{xcolor}

\newcommand*\aap{A\&A}

\newcommand*\apj{ApJ}
\newcommand*\apjl{ApJ}

\newcommand*\jgr{J.~Geophys.~Res.}

\newcommand*\mnras{MNRAS}

\newcommand*\ssr{Space~Sci.~Rev.}

\titlehead{Research}

\begin{document}

\title{On the Ambipolar Diffusion Formulation for Ion-neutral drifts in the non-negligible drift velocity limit}

\author{
A. S. Hillier$^{1}$}

\address{$^{1}$Department of Mathematics and Statistics, University of Exeter, Exeter EX4 4QF U.K.}

\subject{xxxxx, xxxxx, xxxx}

\keywords{xxxx, xxxx, xxxx}

\corres{A. Hillier\\
\email{a.s.hillier@exeter.ac.uk}}

\begin{abstract}
The ambipolar diffusion approximation is used to model partially ionised plasma dynamics in a single fluid setting. 
To correctly apply the commonly used version of ambipolar diffusion, a set of criteria should be satisfied including the requirement that the difference in velocity between charges and neutral species (known as drift velocity) is much smaller than the thermal velocity, otherwise the drift velocity will drive a non-negligible level of further collisions between the two species. 
In this paper we explore the consequence of relaxing this assumption. We show that a new induction equation can be formulated without this assumption. This formulation reduces to the ambipolar induction equation in the case the drift velocity is small. In the large drift velocity limit, the feedback of the drift velocity on the collision frequency results in decreased diffusion of the magnetic field compared with the standard ambipolar diffusion approximation for the same parameters. This has a natural consequence of reducing the frictional heating that can occur. Applying this to results from flux emergence simulations where the expansion of the magnetic field leads to strong adiabatic cooling of the partially ionised chromosphere resulted in a noticeable reduction in the magnitude of the predicted drift velocities.

\end{abstract}






\maketitle

\section{Introduction}

In many astrophysical systems (e.g. stellar atmospheres, jets from young stellar objects and protoplanetary disks), the conditions are such that there is insufficient energy to fully ionise the plasma. Therefore, the majority of the fluid is composed of neutral species. So even though magnetic fields are recognised as being crucial in driving dynamics in these systems, under these conditions, most of the species do not directly feel the Lorentz force of the magnetic field. {As a consequence, charged and neutral species feel different forces. Consequently, this drives a drift in the velocity between these different species} \cite{KHOM2014, KHOM2017, Khomenko2020, MART2015}.
These physical processes are found to be important in a wide range of astrophysical plasmas \cite{BALLE2018}.

This drift between the charged and neutral species means that when the particles collide, momentum is transferred between the plasma and neutral fluids.
Over time, this results in the momentum being redistributed such that the two fluids move together.
The collisions of the particles as they drift past each other can result in frictional heating, meaning that coupling process can be dynamically important for energy dissipation \cite{KHOM2012}.

In situations where the dynamics of the neutrals and the ions happens at sufficiently high frequency the neutrals and charges can act as separate fluids, an example of this will be in the rapid transitions of physical properties found in two-fluid shock fronts \cite{HILL2016}.
However, in many cases the fluid is sufficiently coupled to be treated approximately as one fluid, but with the drift between species taken into account via a modified Ohm's law.
In this situation the induction equation becomes \cite{BRAG1965, Khomenko2020}:
\begin{equation}
\frac{\partial \mathbf{B}}{\partial t}=\nabla\times\left(\mathbf{v\times B}+ \eta_{\rm AMB}\frac{(\mathbf{J}\times\mathbf{B})\times\mathbf{B}}{{B}^2} \right).
\end{equation}

There are a number of approximations that go into this diffusion approximation, key ones are \cite{BRAG1965}:
\begin{itemize}
\item The changes in the drift velocity occur at low frequency.

\item The temperature of the neutrals and the plasma is the same

\item The magnitude of the drift velocity is much smaller than the thermal velocity.
\end{itemize}
Of these approximations, it is worth highlighting the final in this list. This is an often forgotten part of the approximations made when calculating the collision frequency. Investigating the relaxation of this assumption is the key goal of this paper. 

In a low-$\beta$ regime (where plasma $\beta$ is the ratio of gas to magnetic pressure) it is possible there will develop a strong Lorentz force that drives a large enough drift velocity that the associated speed becomes greater than the plasma thermal velocity.
Large velocity drifts, evidenced by the magnitude of the ambipolar diffusion term, have been observed in simulations of magnetic flux emergence \cite{LEAKE2006, ARB2007}. These calculations have shown that the inclusion of ambipolar diffusion drastically changed the rate at which the magnetic flux emerged through the solar chromosphere due to the significant magnitude of the diffusion. 
Similarly large ambipolar terms are found in the chromosphere modelling simulations \cite{MART2016, MART2017a, MART2017b}. In these calculations large velocity drifts were found to develop in the partially ionised chromosphere, including in relation to magnetic flux emergence and spicule dynamics. With these large velocity drifts significant heating was found.
However, when ambipolar diffusion gets large, and with it large velocity drifts are present in simulations, there is the possibility that at least the third of the assumptions behind the ambipolar diffusion approximation is invalidated.

To go beyond the first two of the listed approximations requires multi-fluid modelling. For the final of these approximations, we can include the role of super-thermal drift into the ambipolar diffusion approximation. In this paper, we take advantage of the very accurate approximation that has been developed for the collision frequency \cite{DRAINE1986, ZANK2018} and change in collision cross-section \cite{BARNETT1990} to model the large drift velocity limit to present an extension to the ambipolar diffusion term that allows it to hold for drift velocities greater than the thermal velocity.

\section{Determining the collisional velocity}

Our first step is to look at the two-fluid momentum equations, so that we can then simplify this into a single fluid model.
These are
\begin{align}
\rho_n\frac{D \mathbf{v}_{\rm n}}{Dt}=-\nabla p_{\rm n} -\rho_n\mathbf{g} - \alpha(\mathbf{v}_{\rm D},T_{\rm n},T_{\rm p})\rho_{\rm n}\rho_{\rm p}\mathbf{v}_{\rm D},\label{vneqn}\\
\rho_p\frac{D \mathbf{v}_{\rm p}}{Dt}=-\nabla p_{\rm p}+\frac{1}{c}\mathbf{J}\times\mathbf{B} -\rho_p\mathbf{g}+ \alpha(\mathbf{v}_{\rm D},T_{\rm n},T_{\rm p})\rho_{\rm n}\rho_{\rm p}\mathbf{v}_{\rm D},\label{vpeqn}
\end{align}
where $\mathbf{v}_{\rm D}=\mathbf{v}_{\rm n}-\mathbf{v}_{\rm p}$.
By dividing Equation \ref{vneqn} by $\rho_n$ and Equation \ref{vpeqn} by $\rho_p$ then subtracting the latter from the former approximately leads to this equation
\begin{equation}
\frac{D \mathbf{v}_{\rm D}}{Dt}=\frac{1}{\rho_p}\nabla p_{\rm p}-\frac{1}{c\rho_p}\mathbf{J}\times\mathbf{B}-\frac{1}{\rho_n}\nabla p_{\rm n} - \alpha({v}_{\rm D},T_{\rm n},T_{\rm p})(\rho_{\rm n}+\rho_{\rm p})\mathbf{v}_{\rm D}.
\end{equation}

The parameter $\alpha$ when multiplied by a density gives a collision frequency, where these collision frequencies can either be calculated from hard-sphere collisions between particles or through effective collisions driven by charge exchange. 
For these two different effects, $\alpha$ takes slightly different forms.
When looking only at hard-sphere collisions, $\alpha$ is given by \cite{DRAINE1986, OLIVER2016}
\begin{equation}\label{alpha_approx}
\alpha\approx \frac{1}{m_{\rm n}+m_{\rm p}}\frac{8}{3}\sqrt{\frac{2k_{\rm B}T_{\rm r}}{\pi m_{r}}}\left( 1+\frac{9\pi s^2}{64}\right)^{1/2} \Sigma_{\rm pn}({v}_{\rm D},T_{\rm n},T_{\rm p}),
\end{equation}
where
\begin{align}
T_{\rm r}=&\frac{m_{\rm p}T_{\rm n}+m_{\rm n}T_{\rm p}}{m_{\rm n}+m_{\rm p}},\\
m_{\rm r}=&\frac{m_{\rm n}m_{\rm p}}{m_{\rm n}+m_{\rm p}},\\
s=&v_{\rm D}\left(\frac{2k_{\rm B}T_{\rm r}}{ m_{r}}\right)^{-1/2},\label{s_eqn}\\
v_{\rm D}=&|\mathbf{v}_{\rm D}|=|\mathbf{v}_{\rm n}-\mathbf{v}_{\rm p}|.
\end{align}
{Here $\Sigma_{\rm pn}({v}_{\rm D},T_{\rm n},T_{\rm p})$ is the collision velocity dependent cross-section (see, for example, \cite{VRAN2013} for details on the collision velocity dependence) with $\Sigma_{\rm pn}=\Sigma_{\rm np}$. }
If one is only interested in a hydrogen plasma, as we use in the investigations later in this work, we can take $m_{\rm n}=m_{\rm p}=m_{\rm i}$, where $m_{\rm i}$ is the mass of a proton.

\begin{figure}[ht]
\centering
\includegraphics[height=7cm]{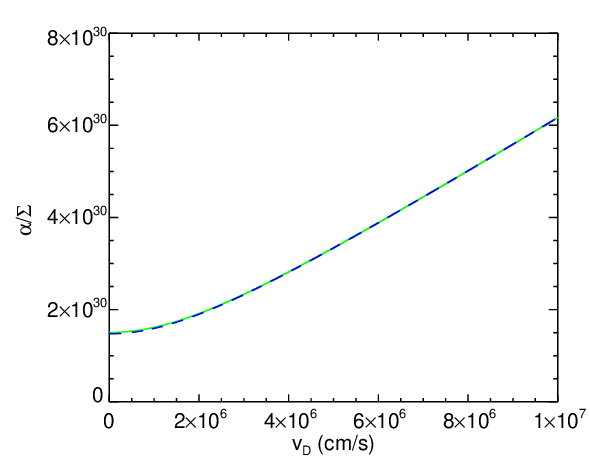}
\caption{Plot of $\alpha/\Sigma$ {(in units of cm/s) against ${v}_{\rm D}$ with $T=8000$\,K} for the approximation shown in equation \ref{alpha_approx} (dashed blue line) and for the model applied in \cite{PAULS1995, MEIER2012, LEAKE2012, LEAKE2013, POP2019}  for charge exchange (solid green line).}
\label{alpha}
\end{figure}

For charge exchange (in this case for hydrogen proton charge exchange) {an equivalent form is used \cite{ZANK2018}
\begin{equation}\label{alpha_approx2}
\alpha\approx \frac{1}{m_{\rm p}}\frac{8}{3}\sqrt{\frac{2k_{\rm B}T_{\rm r}}{\pi m_{r}}}\left( 1+\frac{9\pi s^2}{64}\right)^{1/2} \Sigma_{\rm CX}({v}_{\rm D},T_{\rm n},T_{\rm p}).
\end{equation}
For the two different equations for $\alpha$, there are two key differences. 
Firstly, this is that the cross section for hard sphere collisions and charge exchange are different. }
Secondly, the factor ${1}/(m_{\rm n}+m_{\rm p})$ in Equation \ref{alpha_approx} becomes ${1}/{m_{\rm p}}$ in Equation \ref{alpha_approx2}.
Note that in the literature there are other forms used for $\alpha$ when charge exchange is considered \cite{PAULS1995, MEIER2012, LEAKE2012, LEAKE2013, POP2019}.
As shown in Figure \ref{alpha}, the approximation used in their works (solid green line), and the one adopted in ours (dashed blue line) gives to very high level of accuracy the same dependence of $\alpha/\Sigma$ for different drift velocities {with at most $2$\% difference between the two models}).

\section{Calculating $\mathbf{v}_{\rm D}$ in the strong coupling limit}

In this section, for simplicity we assume that the {cross-section is independent of the drift velocity (many models take the collisional cross-section to be a constant for example). 
We will look later at how the requirement for the cross-section to independent of the drift velocity} can be relaxed. Here we will use the form of $\alpha$ suitable for hard sphere collisions, but it is trivial to rework these derivations to take charge exchange as the principle momentum transfer process into account.

The key assumption of ambipolar diffusion comes from the strong coupling limit which implies that the time derivative of the drift velocity is negligible and that the temperatures of the fluids must all be the same. By assuming, for simplicity of the presentation of the mathematics, that we are dealing with a hydrogen fluid (e.g. $m_{\rm p}=m_{\rm n}$), this leads to:
\begin{equation}\label{basic}
\rho_{\rm T}\frac{1}{2m_{\rm p}}\frac{8}{3\sqrt{\pi}}\sqrt{\frac{4k_{\rm B}T}{m_{p}}}\left( 1+\frac{9\pi s^2}{64}\right)^{1/2} \Sigma_{\rm pn}(T)\mathbf{v}_{\rm D}=-\frac{\mathbf{F}_{\rm p}}{\rho_p}+\frac{\mathbf{F}_{\rm n}}{\rho_n},
\end{equation}
where we have taken (as is usual in the ambipolar diffusion formulation) that $T_{\rm p}=T_{\rm n}=T_{\rm r}=T$, $\mathbf{F}_{\rm n}$ and $\mathbf{F}_{\rm p}$ are denoting the force terms from the RHS of Equations \ref{vneqn} and \ref{vpeqn} respectively, and taking that $\rho_{\rm T}\equiv \rho_{\rm p}+\rho_{\rm n}$.
This can be written as
\begin{equation}\label{basic_2}
\left( 1+\frac{9\pi s^2}{64}\right)^{1/2}s\hat{\mathbf{v}}_{\rm D}=-X\hat{\mathbf{F}}_{\rho},
\end{equation}
where 
\begin{equation}
\mathbf{F}_{\rho}\equiv \frac{1}{\rho_p}\mathbf{F}_{\rm p}-\frac{1}{\rho_n}\mathbf{F}_{\rm n}
\end{equation}
and with
\begin{equation}
X=\frac{2m_{\rm p}}{\rho_{\rm T}}\frac{3\sqrt{\pi}}{8}\left({\frac{4k_{\rm B}T}{m_{p}}}\right)^{-1} \Sigma_{\rm pn}(T)^{-1}|\mathbf{F}_{\rho}|,
\end{equation}
and the hat symbol denotes a unit vector implying
\begin{equation}
\hat{\mathbf{v}}_{\rm D}=-\hat{\mathbf{F}}_{\rho}.
\end{equation}

Squaring both sides of Equation \ref{basic_2}, we can write the following quadratic in $s^2$
\begin{equation}
\frac{9\pi}{64}s^4+s^2-X^2=0.
\end{equation}
This gives the following solution for $s^2$ (noting that $s^2\ge 0$)
\begin{equation}
s^2=\frac{32}{9\pi}\left(\sqrt{1+\frac{9\pi}{16}X^2}-1\right).
\end{equation}
Therefore, $s$ is given by
\begin{equation}\label{s_eqn_1}
s=\sqrt{\frac{32}{9\pi}\left(\sqrt{1+\frac{9\pi}{16}X^2}-1\right)}.
\end{equation}
As such, our velocity drift is given by
\begin{equation}\label{the_drift}
\mathbf{v}_{\rm D}=-\sqrt{\frac{32}{9\pi}\left(\sqrt{1+\frac{9\pi}{16}X^2}-1\right)}\sqrt{\frac{4k_{\rm B}T}{m_{\rm p}}}\hat{\mathbf{F}}_{\rho}.
\end{equation}

\begin{figure}[ht]
\centering
\includegraphics[height=8cm]{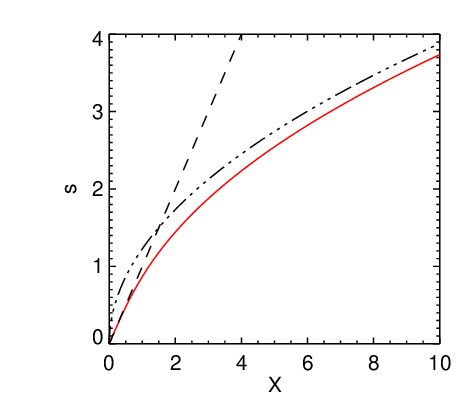}
\caption{Plot of $s$ for a {drift-velocity} independent cross section against $X$ (red line) {as given by Equation \ref{s_eqn_1}}. The two limits, small $X$ limit (dashed line) {with $s=X$ which is equivalent to the standard ambipolar diffusion approximation } and large $X$ limit (dash-triple dot line), are also plotted. }
\label{vd_fixed}
\end{figure}

Figure \ref{vd_fixed} shows the variation of $s$ against $X$ (red line). The dashed line shows the curve expected for $s$ when $X^2$ is taken to be small, and the triple-dot-dashed line shows the curve for $s$ when $X^2$ is assumed to be large. {The derivation of these two limits is explained in the rest of the subsection.} It is clear that as $X$ becomes greater than 1 the value of $s$ strongly diverges from that of the small-force limit.

\subsection{The Ambipolar diffusion approximation in the Lorentz force dominant limit}

Generally for ambipolar diffusion in many astrophysical systems, the largest term on the RHS of Equation \ref{basic} is the one given by the Lorentz force. In this limit the drift velocity is given by:
\begin{equation}\label{LF_drift_eqn}
\mathbf{v}_{\rm D}=-v_{\rm D}\frac{\mathbf{J}\times\mathbf{B}}{|\mathbf{J}\times\mathbf{B}|}.
\end{equation}
Using this drift velocity the induction equation can be rewritten in terms of the neutral velocity
\begin{align}\label{induction}
\frac{\partial \mathbf{B}}{\partial t}=&\nabla\times(\mathbf{v}_{\rm p}\times \mathbf{B})=\nabla\times(\mathbf{v}_{\rm n}\times \mathbf{B}-\mathbf{v}_{\rm D}\times \mathbf{B})
\\=&\nabla\times\left(\mathbf{v}_{\rm n}\times \mathbf{B}\right)+\nabla\times\left(\sqrt{\frac{32}{9\pi}\left(\sqrt{1+\frac{9\pi}{16}X^2}-1\right)}\sqrt{\frac{4k_{\rm B}T}{m_{\rm p}}}\left(\frac{\mathbf{J}\times\mathbf{B}}{|\mathbf{J}\times\mathbf{B}|}\right)\times \mathbf{B}\right).
\end{align}
An alternative way of formulating these equations can be to put the system into the centre-of-mass reference frame \cite{Khomenko2020} which gives
\begin{align}\label{induction2}
\frac{\partial \mathbf{B}}{\partial t}=&\nabla\times(\mathbf{v}_{\rm p}\times \mathbf{B})=\nabla\times(\mathbf{v}_{\rm CM}\times \mathbf{B}-\xi_{\rm n}\mathbf{v}_{\rm D}\times \mathbf{B})
\\=&\nabla\times\left(\mathbf{v}_{\rm CM}\times \mathbf{B}\right)+\nabla\times\left(\xi_{\rm n}\sqrt{\frac{32}{9\pi}\left(\sqrt{1+\frac{9\pi}{16}X^2}-1\right)}\sqrt{\frac{4k_{\rm B}T}{m_{\rm p}}}\left(\frac{\mathbf{J}\times\mathbf{B}}{|\mathbf{J}\times\mathbf{B}|}\right)\times \mathbf{B}\right).
\end{align}
For the rest of the paper we will focus on the neutral frame-of-reference, but as can be seen everything we present here can be easily adapted for systems where the centre-of-mass frame of reference is more appropriate.

\subsubsection{Small-force limit}

To look into the behaviour of the drift velocity (and how it connects to formulations of the induction equation, and the evolution of the magnetic field this allows), it is informative to look at the formulation for $v_{\rm D}$ in the limits of small and large Lorentz force. 

The first limit we investigate is the limit of small forces, which can be described as the situation where
\begin{equation}
1\gg \frac{9\pi}{16}X^2.
\end{equation}
In this limit, we can perform a Taylor expansion{, truncating at first order,} and on substitution into Equation \ref{LF_drift_eqn} we find:
\begin{align}
\mathbf{v}_{\rm D}=&-X\left({\frac{4k_{\rm B}T}{m_{\rm p}}}\right)^{1/2}\frac{\mathbf{J}\times\mathbf{B}}{|\mathbf{J}\times\mathbf{B}|},\label{the_drift_small}\\
=&-\frac{2m_{\rm p}}{\rho_{\rm T}^2\xi_{\rm p} \Sigma_{\rm pn}(T)}\frac{3\sqrt{\pi}}{8}\left({\frac{4k_{\rm B}T}{m_{\rm p}}}\right)^{-1/2} \frac{1}{c}\mathbf{J}\times\mathbf{B},
\end{align}
{which is equivalent to $s=X$.}
This is the formulation we would have expected if we had made the assumption of small drift velocity in the original derivation.
In this case Equation \ref{induction} becomes:
\begin{align}
\frac{\partial \mathbf{B}}{\partial t}=&\nabla\times\left(\mathbf{v}_{\rm n}\times \mathbf{B}+\frac{2m_{\rm p}}{\rho_{\rm T}^2\xi_{\rm p} \Sigma_{\rm pn}(T)}\frac{3\sqrt{\pi}}{8}\left({\frac{4k_{\rm B}T}{m_{\rm p}}}\right)^{-1/2} \frac{1}{c}(\mathbf{J}\times\mathbf{B})\times \mathbf{B}\right),
\\=&\nabla\times\left(\mathbf{v}_{\rm n}\times \mathbf{B}+\eta_{\rm AMB}\frac{(\mathbf{J}\times\mathbf{B})\times \mathbf{B}}{B^2}\right),
\end{align}

\subsubsection{Large-force limit}

An alternative limit to investigate, and one that takes our analysis into a regime that is not accessible to standard ambipolar diffusion scenarios, is the limit where the forces on the system are large, which implies we can have large velocity drifts. 
We can simply state this as the regime where
\begin{equation}
1\ll \sqrt{\frac{9\pi}{16}}X.
\end{equation}
In this case
\begin{align}
\mathbf{v}_{\rm D}=&-\sqrt{\frac{ 4kT}{\pi m_{\rm p}}}\sqrt{\frac{8X}{3\sqrt{\pi}}}\frac{\mathbf{J}\times\mathbf{B}}{|\mathbf{J}\times\mathbf{B}|},
\\=&-\sqrt{\frac{2m_{\rm p}}{\pi\rho_{\rm T}^2\xi_{\rm p}\Sigma_{\rm pn}(T)}}\frac{1}{\sqrt{c}}\frac{\mathbf{J}\times\mathbf{B}}{\sqrt{|\mathbf{J}\times\mathbf{B}|}},
\end{align}
which again can be derived from Equation \ref{basic} {in this case setting $X\gg 1$ in Eqn \ref{s_eqn_1} resulting in $s=\sqrt{8X/3\sqrt{\pi}}$}.
As we can clearly see, in this limit the velocity drift scales as $\sqrt{|\mathbf{J}\times\mathbf{B}|}$ instead of $|{\mathbf{J}\times\mathbf{B}|}$ for the small-force limit.
In this limit, the induction equation becomes
\begin{align}
\frac{\partial \mathbf{B}}{\partial t}=&\nabla\times\left(\mathbf{v}_{\rm n}\times \mathbf{B}+\sqrt{\frac{2m_{\rm p}}{\pi\rho_{\rm T}^2\xi_{\rm p}\Sigma_{\rm pn}(T)}}\frac{1}{\sqrt{c}}\frac{\mathbf{J}\times\mathbf{B}}{\sqrt{|\mathbf{J}\times\mathbf{B}|}}\times \mathbf{B}\right).
\end{align}

\subsection{Consequences for ambipolar diffusion heating}

Based on {the summation of} the two-fluid energy internal energy equations \cite{Khomenko2020}, the change in the total internal energy {(as relevant for a single-fluid energy equation)} as a result of frictional heating ($H$) is given by:
\begin{align}
H=&\alpha(\mathbf{v}_{\rm D},T)\rho_{\rm T}^2\xi_{\rm p}\xi_{\rm n}{v}_{\rm D}^2,\\
=& \frac{\rho_{\rm T}^2\xi_{\rm n}\xi_{\rm p}}{m_{\rm p}}\frac{128}{27}\left(\frac{4k_{\rm B}T}{\pi m_{\rm p}}\right)^{3/2}\Sigma_{\rm pn}(T)\frac{1}{\sqrt{2}}\left(1+\sqrt{1+\frac{9\pi X^2}{16}} \right)^{1/2}\left(\sqrt{1+\frac{9\pi X^2}{16}}-1 \right).
\end{align} 
where
\begin{equation}
\frac{1}{\gamma-1}\frac{\partial p}{\partial t}=H,
\end{equation}
when there is only frictional heating in the system and $p=p_{\rm n}+p_{\rm p}$.
If we take
\begin{equation}
Y(m_{\rm p},\rho_{\rm T},\xi_{\rm n},\xi_{\rm p},T)=\frac{\rho_{\rm T}^2\xi_{\rm n}\xi_{\rm p}}{m_{\rm p}}\frac{128}{27}\left(\frac{4k_{\rm B}T}{\pi m_{\rm p}}\right)^{3/2}\Sigma_{\rm pn}(T),
\end{equation}
then we can write this as
\begin{align}
H=&\frac{1}{\sqrt{2}}Y\left(1+\sqrt{1+\frac{9\pi X^2}{16}} \right)^{1/2}\left(\sqrt{1+\frac{9\pi X^2}{16}}-1 \right).
\end{align}


Again we can look at the different limits, first taking the limit of small $X^2$. This gives:
\begin{align}
H=& Y\frac{9\pi}{32}X^2\\
=& \frac{m_{\rm p}\xi_n}{\rho_{\rm T}^2\xi_{\rm p}}\frac{3\sqrt{\pi}}{4}\left(\frac{4k_{\rm B}T}{m_{\rm p}}\right)^{-1/2}\Sigma_{pn}^{-1}\left|\frac{1}{c}\mathbf{J\times B}\right|^2
\end{align} 
Alternatively, in the large force limit
\begin{align}
H=& \frac{Y}{\sqrt{2}}\left(\frac{9\pi}{16}\right)^{3/4}X^{3/2}\\
=& 4\left(\frac{3}{2}\right)^{3/2}\xi_{\rm n}\left(\frac{m_{\rm p}}{\rho_{\rm T}^2\xi_{\rm p}}\right)^{1/2}\Sigma_{pn}^{-1/2}\left|\frac{1}{c}\mathbf{J\times B}\right|^{3/2}.
\end{align} 
Therefore, when the Lorentz force in the system is small the heating scales with $|\mathbf{J}\times\mathbf{B}|^2$. However, for large Lorentz force the heating then scales as $|\mathbf{J}\times\mathbf{B}|^{3/2}$. Though not as extreme a difference as the velocity difference scaling (and with it the magnitude of ambipolar diffusion) with $|\mathbf{J}\times\mathbf{B}|$ changing to $|\mathbf{J}\times\mathbf{B}|^{1/2}$. But this still illustrates an important consequence, in low-$\beta$ systems the {heating rates that are calculated using the standard ambipolar diffusion approximation for single fluid models (i.e. the small force limit of this paper) can easily be an overestimate if the role of velocity drift in driving collisions is neglected}.



\section{Beyond singles species and constant cross-sections}

The previous example looked at a fluid with only a single species (nominally hydrogen but the analysis would apply to any species), and where the collisional cross-section was independent of velocity drift. In this section we present some extensions to this model.

\subsection{Investigating the role of a non-constant collisional cross-section}

\begin{figure}[ht]
\centering
\includegraphics[height=7cm]{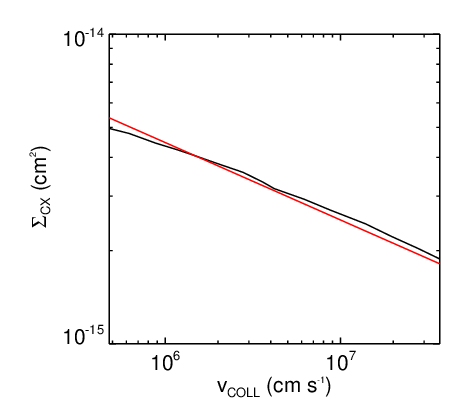}
\caption{Plot of $\Sigma_{\rm CX}$ against $v_{\rm COLL}$. The black line gives the experimental data from \cite{BARNETT1990} and the red line represents the fit used in this study.}
\label{cross-sec}
\end{figure}

Until now, we have worked on the assumption that the collisional cross-section is independent of the drift velocity.
However, this is not going to be the case.
For example, in the case of charge exchange, the cross section for neutral-proton interactions can be approximated by the form \cite{MEIER2012}
\begin{equation}
\Sigma_{\rm CX}=1.12\times 10^{-14}-7.15\times 10^{-16}\ln(v_{\rm COLL}/100) {\,\mathrm cm}^2,
\end{equation}
with
\begin{equation}
v_{\rm COLL}=\sqrt{\frac{16 k T}{\pi m_p}+v_{\rm D}^2},
\end{equation}
note this form for the cross section is taken as a fit to data \cite{BARNETT1990} {with the accuracy of the fit estimated as 10\%}.
Looking at this form, it is clearly awkward to manipulate algebraically, so if we return to the data for the charge exchange cross section \cite{BARNETT1990}, by looking at their figure we can see that then the magnitude of the cross section looks as if it varies as an approximate power law over a large range of the data.
By fitting the data of \cite{BARNETT1990} in the range of $v_{\rm COLL}=4.79\times 10^5$ to $3.68\times 10^7$\,cm\,s$^{-1}$ with a function of the form $\Sigma_{\rm CX}=A v_{\rm COLL}^{-1/n}$ with $A$ a positive constant $n$ a positive integer, we attempt to approximate the change in charge exchange cross section as a function of velocity.
We found:
\begin{equation}
\Sigma_{\rm CX}=10^{-12.85}v_{\rm COLL}^{-1/4} {\,\mathrm{cm}}^2=\Sigma_{\rm CX,0}(T)\left(1+\frac{\pi}{4}s^2\right)^{-1/8},
\end{equation}
with $\Sigma_{\rm CX,0}(T)$ the cross section at zero drift velocity.
Note that this fit is within 5\% of the atomic data. The accuracy can be seen in Figure \ref{cross-sec} (and has the benefit over that of \cite{MEIER2012} of being in the form of a power law and as such allows easier algebraic manipulation). Due to the restrictions in fitting, more accurate fits are possible, but the accuracy of the fit used here is more than sufficient for the work.
It is important to note that more accurate models of the cross section exist, for example those published in \cite{VRAN2013}, where quantum mechanical effects are taken into account resulting in non-monotonic changes in the cross-section at energies below 1\,ev. However, we believe that the cross-section used here is sufficiently accurate for our purposes.

We can rewrite Equation \ref{basic} (assuming the only force is the Lorentz force) but including a {drift-velocity} dependent cross-section giving
\begin{equation}
\frac{\rho_{\rm T}}{m_{\rm p}}\frac{8}{3\sqrt{\pi}}\sqrt{\frac{4k_{\rm B}T}{ m_{p}}}\left( 1+\frac{9\pi s^2}{64}\right)^{1/2} \Sigma_{\rm CX,0}(T)\left(1+\frac{\pi}{4}s^2\right)^{-1/8}\mathbf{v}_{\rm D}=-\frac{1}{\rho_{\rm p}}\mathbf{J\times B},
\end{equation}
or
\begin{equation}
\left( 1+\frac{9\pi s^2}{64}\right)^{1/2}\left(1+\frac{\pi}{4}s^2\right)^{-1/8}s\hat{\mathbf{v}}_{\rm D}=-X_{\rm CX}\frac{\mathbf{J\times B}}{|\mathbf{J\times B}|},
\end{equation}
with $X_{\rm CX}$ defined as
\begin{equation}
X_{\rm CX}=\frac{m_{\rm p}}{\rho_{\rm T}^2\xi_{\rm p}}\frac{3\sqrt{\pi}}{8}\left({\frac{4k_{\rm B}T}{ m_{p}}}\right)^{-1/2} \Sigma_{\rm CX,0}(T)^{-1}\left|\frac{\mathbf{J\times B}}{c}\right|.
\end{equation}
By rearranging and putting to the 8$^{\rm th}$ power we find
\begin{equation}\label{cx_solution}
\left( 1+\frac{9\pi s^2}{64}\right)^{4}s^8=X_{\rm CX}^8\left(1+\frac{\pi}{4}s^2\right).
\end{equation}

\begin{figure}[ht]
\centering
\includegraphics[height=8cm]{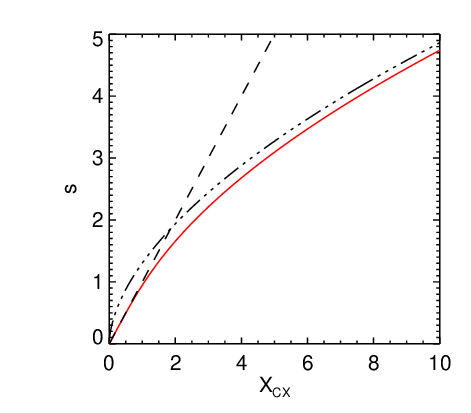}
\caption{Plot of $s$ for a {drift-velocity} dependent cross section against $X_{\rm CX}$ (red line) {calculated by taking the appropriate solution of Equation \ref{cx_solution}}. The two limits, small $s$ limit (dashed line) and large $s$ limit (dash-triple dot line), are also plotted.}
\label{vd_2}
\end{figure}

Rearranging Equation \ref{cx_solution} for $s$ gives an 8$^{\rm th}$ order polynomial in $s^2$. The roots of this equation can be solved numerically, and then taking that we require $s^2$ to be both purely real and positive to give physical solutions we can find the appropriate solution and then determine $s$.
We can also look again at the limits, in this case when $s$ is either small or large.
For $s^2\ll 1$ we have
\begin{equation}
s=X_{\rm CX}.
\end{equation}
Alternatively, for $s^2\gg 1$ we have
\begin{equation}
s=X_{\rm CX}^{4/7}\left({\frac{64}{9\pi}}\right)^{2/7}\left( \frac{\pi}{4}\right)^{1/14}.
\end{equation}
Therefore, unlike the constant cross section case, we have the dependence of $v_{\rm D}$ on $|\mathbf{J\times B}|$ as scaling as $|\mathbf{J\times B}|^{4/7}$ in the large drift velocity limit (equivalent to the large force limit used earlier) instead of $|\mathbf{J\times B}|^{1/2}$ found in the case of a constant cross section.
Note this implies the scaling of the heating rate with the magnitude of the Lorentz force of $|\mathbf{J\times B}|^{12/7}$.
Figure \ref{vd_2} shows $s$ against $X_{\rm CX}$, the red line is the solution for $s$, and the dashed and triple dot dash lines are the small and large $s$ solutions respectively.

\section{Summary and Conclusion}

In this paper, we have provided a new analytic framework to relax a key assumption behind the ambipolar diffusion formulation, something which is used to model the influence of neutrals in the MHD induction equation. Namely, we have removed the requirement that the difference in velocities between the charged and neutral fluids has to be much less than the thermal velocity of the plasma. The key consequence of removing this assumption is that it shows that using the standard ambipolar diffusion approximation may overestimate the magnitude of the diffusion of the magnetic field and the heating this produces.

A key question remains, and this is: will there ever be a situation in the solar atmosphere where the velocity drift gets large enough that the newly proposed formulation is required? To date, the measured drift velocities, though of order 1\,km\,s$^{-1}$ (for example \cite{ZAPIOR2022}), have all been too small to provide a meaningful departure from the ambipolar diffusion coefficient predicted by the small drift velocity limit. However, simulations have shown that there is the potential for huge velocity drifts. For example in the simulations (including the standard ambipolar diffusion approximation) of emerging flux  by \cite{NOBREGA2020}, they found drift velocities of $\sim 40$\,km\,s$^{-1}$ in regions where the temperature had dropped to $\sim 2,000$\,K. Putting these parameters into Equation \ref{s_eqn} gives a value of $s\approx 5$, which from Figure \ref{vd_fixed} we can see corresponds to a value of $X=5$ {as in this limit $s=X$}. Taking into account that this large drift velocity would drive further particle collisions acting as a limiter to the ambipolar diffusion we find a more accurate value of $s\approx 2.5$ (equivalent to a drift velocity of $\sim 20$\,km\,s$^{-1}$). Though this drift velocity is still very significant, it represents a non-trivial reduction in magnitude.    

It is reasonable to expect that a detailed investigation of the output of single-fluid partially ionised plasma models of the solar atmosphere would highlight examples where the changes in magnitude of the diffusion were even larger. 
For example, in two-fluid simulations of magnetic reconnection \cite{MURTAS2021, MURTAS2022} it has been necessary to use this modified form of $\alpha$ as the effect of drift-induced collisions became significant, so an important future work is to assess during which physical processes the assumption of small drift velocity breaks down. Beyond the solar atmosphere, there are many astrophysical systems with low temperatures and densities. If this is combined with being low plasma $\beta$, then the modified formulation presented here may be applicable.  


\ack{AH was supported in this research by his STFC Ernest Rutherford Fellowship grant number ST/L00397X/2 and STFC research grants ST/R000891/1 and ST/V000659/1. AH would like to acknowledge the discussions with members of ISSI Team 457 ``The Role of Partial Ionization in the Formation, Dynamics and Stability of Solar Prominences'', which have helped improve the ideas in this manuscript.}

\bibliographystyle{RS} 

\end{document}